\documentclass[12pt]{article}
\usepackage{amsmath,amssymb,bm,graphicx}
\usepackage{color} 

\newcommand{\delslash}{\not \! \partial}

\begin{document}


\begin{center}
{\Large{\bf Generalized Pauli-Gursey transformation and Majorana neutrinos}}
\end{center}
\vskip .5 truecm
\begin{center}
{\bf { Kazuo Fujikawa}}
\end{center}

\begin{center}
\vspace*{0.4cm} 
{\it {Interdisciplinary Theoretical and Mathematical Sciences Program,\\
RIKEN, Wako 351-0198, Japan
}}
\end{center}
\vspace*{0.2cm} 
\begin{abstract}
We discuss a generalization of the Pauli-Gursey transformation, which is motivated by the Autonne-Takagi factorization, to an arbitrary $n$ number of generations of neutrinos using  $U(2n)$ that defines general canonical transformations and diagonalizes symmetric complex Majorana mass matrices in special cases. The Pauli-Gursey transformation mixes particles and antiparticles and thus changes the definition of the vacuum and C. We define C, P and CP symmetries at each Pauli frame specified by a generalized Pauli-Gursey transformation. The Majorana neutrinos in the C and P violating seesaw model are then naturally defined by a suitable choice of the Pauli frame, where only Dirac-type fermions appear with well-defined  C, P and CP, and thus the C symmetry for Majorana neutrinos agrees  with the C symmetry for Dirac-type fermions. This fully symmetric setting corresponds to the idea of Majorana neutrinos as Bogoliubov quasi-particles. In contrast, the conventional direct construction of Majorana neutrinos in the seesaw model, where CP is well-defined but C and P are violated, encounters the mismatch of C symmetry for Majorana neutrinos and C symmetry for chiral fermions; this mismatch is recognized as the inevitable appearance of the singlet (trivial) representation of C symmetry for chiral fermions.

\end{abstract}
\section{Introduction}
 The symmetry principle or symmetry transformation law in field theory should be defined independently whether some symmetry is a good symmetry of a given Lagrangian or not. For example, the breaking of C, P and CP in weak interactions in the Standard Model is usually analyzed by looking at the breaking of these symmetries which are originally defined as the symmetries of the free part of the Lagrangian. The principle of defining C, P and CP for Majorana fermions starting with chiral fermions is however not very clear, as is explained in further detail later. We discuss this issue by taking the seesaw model Lagrangian for Majorana neutrinos as an example using a generalization of the Pauli-Gursey transformation. In this formulation, it is shown that the Majorana neutrinos are naturally defined in a specific Pauli frame defined by a suitable choice of the generalized Pauli-Gursey transformation despite the fact that the starting seesaw Lagrangian is C and P violating.

We start with a model Lagrangian for the three generations of neutrinos,
\begin{eqnarray}\label{Lagrangian}
{\cal L}&=&\overline{\nu}_{L}(x)i\gamma^{\mu}\partial_{\mu}\nu_{L}(x)+\overline{\nu}_{R}(x)i\gamma^{\mu}\partial_{\mu}\nu_{R}(x)\nonumber\\
&-&\overline{\nu}_{L}(x)m_{D} \nu_{R}(x)
-(1/2)\nu_{L}^{T}(x)C m_{L}\nu_{L}(x)\nonumber\\
&-&(1/2)\nu_{R}^{T}(x)C m_{R}\nu_{R}(x) + h.c.
\end{eqnarray}
where $m_{D}$ is a  $3\times 3$ complex Dirac mass matrix, and $m_{L}$ and $m_{R}$ are $3\times 3$ Majorana mass matrices. 
The anti-symmetry of the matrix $C$, $C^{T}=-C$, and Fermi statistics imply that $m_{L}$ and $m_{R}$ are symmetric and generally complex. We follow the notational conventions in \cite{bjorken}, in particular, $C=i\gamma^{2}\gamma^{0}$.
For $m_L=0$, \eqref{Lagrangian} represents the seesaw Lagrangian of type I~\cite{xing}. In the following, we shall call the expression \eqref{Lagrangian} that covers the main features of the model which generates Majorana neutrinos as the seesaw Lagrangian~\cite{xing, mohapatra1, fukugita, giunti, bilenky,valle}.
 
Before entering on detailed analyses, we would like to briefly state the basic problem we want to solve, its historic background  and a summary of the present analysis.  The Lagrangian \eqref{Lagrangian} is not left-right symmetric and thus the standard definition of P is violated. If one assumes CP invariance, the charge conjugation C is then substantially  broken. On the other hand, the exact solutions of  \eqref{Lagrangian}  are Majorana neutrinos which are the exact eigenstates of C. Apparently, a simple definition of C does not work.
To resolve this puzzing aspect is the main purpose of the present paper. 

The conventional formulation of Majorana neutrinos is  summarized in Section 3 later. The most common formulation of the seesaw scheme  is based on the use of ``pseudo C-symmetry'' defined in \eqref{pseudo C operator}, but 
the modified C is operatorially undefined  as is shown in \eqref{operatorial inconsistency}.  Another possibility in the conventional formulation is to adopt the C symmetry of the emerging Majorana neutrinos as a primary definition of C. It is shown in \eqref{induced-transformations} in Section 3 that this leads to the trivial representation of C for the original chiral fermions in the starting Lagrangian \eqref{Lagrangian}.
This latter possibility is a novel finding of the present paper, but the solution, in our opinion, is not very natural.  

The most natural solution, we believe, is to use a relativistic analogue of Bogoliubov transformation by treating the C violating part of the Lagrangian as a C-condensation, analogously to the charge-nonconserving electron pair term in the BCS theory. This is based on an analogous form we recognize of the seesaw Lagrangian and the BCS theory. This idea leads to “Majorana neutrino as Bogoliubov quasi-particle” we discussed elsewhere. Main part of the present paper in Section 2 is a reformulation of this idea using a generalization of the Pauli-Gursey transformation, which has been discussed only for the single flavor in the past, to multi-flavor cases. The generalized Pauli-Gursey transformation, which contains  the Bogoliubov transformation mentioned above as a special case,  mixes the fermion and antifermion and thus describes a multiple structure of vacua with C symmetry defined for each vacuum. It provides a convenient machinery to formulate Majorana neutrinos from a wider perspective.

\section{Generalized Pauli-Gursey transformation} 
We analyze the Lagrangian \eqref{Lagrangian} by writing the mass term as 
\begin{eqnarray}
(-2){\cal L}_{mass}=
\left(\begin{array}{cc}
            \overline{\nu_{R}}&\overline{\nu_{R}^{C}}
            \end{array}\right)
\left(\begin{array}{cc}
            m_{R}& m_{D}\\
            m^{T}_{D}&m_{L}
            \end{array}\right)
            \left(\begin{array}{c}
            \nu_{L}^{C}\\
            \nu_{L}
            \end{array}\right) +h.c.,
\end{eqnarray}
where we defined 
\begin{eqnarray}
\nu_{L}^{C}\equiv C\overline{\nu_{R}}^T, \ \ \ \nu_{R}^{C}\equiv C\overline{\nu_{L}}^T  
\end{eqnarray}
and used the property
\begin{eqnarray}
\overline{\nu_{R}^{C}}m^{T}_{D}\nu_{L}^{C}&=&\overline{C\overline{\nu_{L}}^T}m^{T}_{D}C\overline{\nu_{R}}^T\nonumber\\
&=&- \nu_{L}^Tm^{T}_{D}\overline{\nu_{R}}^T\nonumber\\
&=&\overline{\nu_{R}}m_{D}\nu_{L}.
\end{eqnarray}
We diagonalize the complex {\em symmetric} mass matrix 
using a $6 \times 6$ unitary matrix (Autonne-Takagi factorization) \footnote{One may start with the bi-unitary transformation 
            $V^{\dagger}
            \left(\begin{array}{cc}
            m_{R}& m_{D}\\
            m^{T}_{D}& m_{L}
            \end{array}\right)
            U
            =\left(\begin{array}{cc}
            M_{1}&0\\
            0&-M_{2}
            \end{array}\right)$ 
            which is written as 
            $U^{T}
            \left(\begin{array}{cc}
            m_{R}& m_{D}\\
            m^{T}_{D}& m_{L}
            \end{array}\right)
            V^{\star}
            =\left(\begin{array}{cc}
            M_{1}&0\\
            0&-M_{2}
            \end{array}\right)$
using the symmetric property of the mass matrix. This implies $V^{\dagger}=U^{T}$ which is shown to be the case with a more detailed analysis~\cite{autonne}.}           
\begin{eqnarray}\label{Autonne-Takagi}
            U^{T}
            \left(\begin{array}{cc}
            m_{R}& m_{D}\\
            m^{T}_{D}& m_{L}
            \end{array}\right)
            U
            =\left(\begin{array}{cc}
            M_{1}&0\\
            0&-M_{2}
            \end{array}\right)    ,        
\end{eqnarray}
where  $M_{1}$ and $M_{2}$ are $3\times 3$ real diagonal matrices, which can be chosen to be the same as characteristic values. Note that we have $U^{T}$ instead of $U^{\dagger}$. We denote one of the eigenvalues as $-M_{2}$ instead of $M_{2}$ by taking into account that we have positive $M_{1},M_{2}=\sqrt{(m_{R}/2)^{2}+m_{D}^{2}}\pm m_{R}/2$, respectively, for the special case of a single generation with real $m_{D}$, $m_{R}$, and $m_{L}=0$.

We thus have
\begin{eqnarray}\label{exact-mass0}
(-2){\cal L}_{mass}
&=& \left(\begin{array}{cc}
            \overline{\tilde{\nu}_{R}}&\overline{\tilde{\nu}_{R}^{C}}
            \end{array}\right)
\left(\begin{array}{cc}
            M_{1}&0\\
            0&-M_{2} 
            \end{array}\right)            
            \left(\begin{array}{c}
            \tilde{\nu}_{L}^{C}\\
            \tilde{\nu}_{L}
            \end{array}\right) +h.c.,                       
\end{eqnarray}
where we defined
\begin{eqnarray} \label{variable-change-0}          
            &&\left(\begin{array}{c}
            \nu_{L}^{C}\\
            \nu_{L}
            \end{array}\right)
            = U \left(\begin{array}{c}
            \tilde{\nu}_{L}^{C}\\
            \tilde{\nu}_{L}
            \end{array}\right)
           ,\ \ \ \ 
            \left(\begin{array}{c}
            \nu_{R}\\
            \nu_{R}^{C}
            \end{array}\right)
            = U^{\star} 
            \left(\begin{array}{c}
            \tilde{\nu}_{R}\\
            \tilde{\nu}_{R}^{C}
            \end{array}\right).          
\end{eqnarray}
Hence we can write 
\begin{eqnarray}\label{exact-solution0}
{\cal L}
&=&(1/2)\{\overline{\tilde{\nu}_{L}}(x)i\delslash\tilde{\nu}_{L}(x)+ \overline{\tilde{\nu}_{L}^{C}}(x)i\delslash \tilde{\nu}_{L}^{C}(x)+\overline{\tilde{\nu}_{R}}(x)i\delslash \tilde{\nu}_{R}(x)\nonumber\\
&& \ \ \ \ \ + \overline{\tilde{\nu}_{R}^{C}}(x)i\delslash \tilde{\nu}_{R}^{C}(x)\}\nonumber\\
&-&(1/2)\left(\begin{array}{cc}
            \overline{\tilde{\nu}_{R}},&\overline{\tilde{\nu}_{R}^{C}}
            \end{array}\right)
\left(\begin{array}{cc}
            M_{1}&0\\
            0&-M_{2} 
            \end{array}\right)            
            \left(\begin{array}{c}
            \tilde{\nu}_{L}^{C}\\
            \tilde{\nu}_{L}
            \end{array}\right) +h.c..
\end{eqnarray}
The fundamental condition, which is essential to define a canonical transformation,
\begin{eqnarray}\label{basic-condition}
\tilde{\nu}_{L}^{C}= C\overline{\tilde{\nu}_{R}}^T, \ \ \ \tilde{\nu}_{R}^{C}= C\overline{\tilde{\nu}_{L}}^T  
\end{eqnarray}
is satisfied after the change of variables \eqref{variable-change-0}, if one notes
\begin{eqnarray} 
\tilde{\nu}_{L}&=&(U^{\dagger})_{21}\nu_{L}^{C} + (U^{\dagger})_{22}\nu_{L},\nonumber\\
 \tilde{\nu}_{R}^{C}&=&(U^{\dagger})^{\star}_{21}\nu_{R}
 +(U^{\dagger})^{\star}_{22}\nu_{R}^{C}
\end{eqnarray} 
 using $3\times 3$ submatrices defined by
\begin{eqnarray}
U^{\dagger}=\left(\begin{array}{cc}
            (U^{\dagger})_{11}&(U^{\dagger})_{12}\\
            (U^{\dagger})_{21}&(U^{\dagger})_{22} 
            \end{array}\right).
\end{eqnarray}
Note that this fundamental condition \eqref{basic-condition} is not generally satisfied by the usual bi-unitary transformation.
The unitary matrix $U$ in \eqref{variable-change-0} transfers all the CP violating effects to the leptonic weak mixing matrix leaving the CP invariant Lagrangian of neutrinos with real diagonal mass parameters. 
We then have
\begin{eqnarray}\label{exact-solution1}
{\cal L}&=&\overline{\tilde{\nu}_{L}}(x)i\delslash \tilde{\nu}_{L}(x)+\overline{\tilde{\nu}_{R}}(x)i\delslash \tilde{\nu}_{R}(x)\nonumber\\
&-&(1/2)\{\tilde{\nu}_{R}^{T}CM_{1}\tilde{\nu}_{R}-\tilde{\nu}_{L}^{T}CM_{2}\tilde{\nu}_{L}\} +h.c..           
\end{eqnarray}
We call these chiral fermions $\tilde{\nu}_{R}(x)$ and $\tilde{\nu}_{L}(x)$, which satisfy the equations of motion
$i\delslash \tilde{\nu}_{R}(x)-M_{1}C\overline{\tilde{\nu}_{R}}^{T}(x)=0$ and 
$i\delslash \tilde{\nu}_{L}(x)+M_{2}C\overline{\tilde{\nu}_{L}}^{T}(x)=0$, respectively, 
chiral mass eigenstates. 

We now introduce classical transformation rules of discrete symmetries. 
We focus our attention on the free kinetic part of the Lagrangian \eqref{Lagrangian}
\begin{eqnarray}\label{free-part1}
{\cal L}_{(0)}&=&\overline{\nu}_{L}(x)i\gamma^{\mu}\partial_{\mu}\nu_{L}(x)+\overline{\nu}_{R}(x)i\gamma^{\mu}\partial_{\mu}\nu_{R}(x)\nonumber\\
&=&\frac{1}{2}\{\overline{\nu}_{L}(x)i\gamma^{\mu}\partial_{\mu}\nu_{L}(x)+\overline{\nu^{C}_{L}}(x)i\gamma^{\mu}\partial_{\mu}\nu^{C}_{L}(x)\nonumber\\
&&+\overline{\nu}_{R}(x)i\gamma^{\mu}\partial_{\mu}\nu_{R}(x)+\overline{\nu^{C}_{R}}(x)i\gamma^{\mu}\partial_{\mu}\nu^{C}_{R}(x)\}
\end{eqnarray}
which determines all the canonical anti-commutation relations, and we treat all the rest of mass terms as interaction terms, analogously to the fact that Heisenberg commutation relations are not modified by (non-derivative)  potential terms in quantum mechanics. We then define C, P and CP, respectively, as the symmetry of this kinetic part of the Lagrangian \cite{bjorken}
\begin{eqnarray}\label{C-P-CP1}
C&:& \nu_{L,R}(x) \rightarrow \nu^{C}_{L,R}(x)=C\overline{\nu_{R,L}}^{T}(x),\nonumber\\
P&:& \nu_{L,R}(x) \rightarrow i\gamma^{0}\nu_{R,L}(t,-\vec{x}),\ \ \nu^{C}_{L,R}(x) \rightarrow i\gamma^{0}\nu^{C}_{R,L}(t,-\vec{x}),\nonumber\\
CP&:& \nu_{L,R}(x)\rightarrow i\gamma^{0}C\overline{\nu_{L,R}(t,-\vec{x})}^{T}
\end{eqnarray}
with a specific definition of "$i\gamma^{0}$ parity" (instead of the conventional "$\gamma^{0}$ parity") explained later; of course, the total Lagrangian is not invariant under this transformation in general. Note that we define the charge conjugation of a Dirac-type fermion by $\psi^{C}(x)=C\overline{\psi}^{T}(x)$, and thus 
$\psi_{R}^{C}(x)=C\overline{\psi_{L}}^{T}(x)$ which implies the C transformation rule $\nu_{R}(x) \rightarrow \nu^{C}_{R}(x)=C\overline{\nu_{L}}^{T}(x)$ and similarly for $\nu_{L}(x)$ by formally treating $\nu_{R}(x)+\nu_{L}(x)$ as a Dirac-type variable.   

We define a transformation, which is motivated by the  Autonne-Takagi factorization \eqref{Autonne-Takagi}  and called the  generalized Pauli-Gursey transformation in this paper for the reasons stated below, using  an {\em arbitrary} $6\times 6$  unitary $U$ by
\begin{eqnarray} \label{variable-change-0-prime}          
            &&\left(\begin{array}{c}
            \nu_{L}^{C}\\
            \nu_{L}
            \end{array}\right)
            = U \left(\begin{array}{c}
            \tilde{\nu}_{L}^{C}\\
            \tilde{\nu}_{L}
            \end{array}\right)
           ,\ \ \ \ 
            \left(\begin{array}{c}
            \nu_{R}\\
            \nu_{R}^{C}
            \end{array}\right)
            = U^{\star} 
            \left(\begin{array}{c}
            \tilde{\nu}_{R}\\
            \tilde{\nu}_{R}^{C}
            \end{array}\right)          
\end{eqnarray}
which still preserves the condition of a canonical transformation as in \eqref{basic-condition}
\begin{eqnarray}\label{basic-condition2}
\tilde{\nu}_{L}^{C}= C\overline{\tilde{\nu}_{R}}^T, \ \ \ \tilde{\nu}_{R}^{C}= C\overline{\tilde{\nu}_{L}}^T. 
\end{eqnarray}
The kinetic part of the Lagrangian after the transformation then becomes
\begin{eqnarray}\label{free-part2}
{\cal L}_{(0)}
&=&(1/2)\{\overline{\tilde{\nu}_{L}}(x)i\delslash\tilde{\nu}_{L}(x)+ \overline{\tilde{\nu}_{L}^{C}}(x)i\delslash \tilde{\nu}_{L}^{C}(x)\nonumber\\
&&+\overline{\tilde{\nu}_{R}}(x)i\delslash \tilde{\nu}_{R}(x) + \overline{\tilde{\nu}_{R}^{C}}(x)i\delslash \tilde{\nu}_{R}^{C}(x)\}
\end{eqnarray}
for which we define C, P and CP transformations, respectively, by 
\begin{eqnarray}\label{C-P-CP2}
C&:& \tilde{\nu}_{L,R}(x) \rightarrow \tilde{\nu}^{C}_{L,R}(x) = C\overline{\tilde{\nu}_{R,L}}^{T}(x),\nonumber\\
P&:& \tilde{\nu}_{L,R}(x) \rightarrow i\gamma^{0}\tilde{\nu}_{R,L}(t,-\vec{x}),\ \ \tilde{\nu}^{C}_{L,R}(x) \rightarrow i\gamma^{0}\tilde{\nu}^{C}_{R,L}(t,-\vec{x}),\nonumber\\
CP&:& \tilde{\nu}_{L,R}(x)\rightarrow i\gamma^{0}C\overline{\tilde{\nu}_{L,R}(t,-\vec{x})}^{T}.
\end{eqnarray}
We note that the Lagrangians \eqref{free-part1} and \eqref{free-part2} clearly show that the generalized Pauli-Gursey transformation with an arbitrary $6\times6$ unitary $U$ defines a canonical transformation by preserving basic anti-commutation relations, since anti-commutation relations are determined by the kinetic part of the Lagrangian independently of mass terms. 
Here an analogy is that $(\nu_{L}(x), \nu_{R}(x))$ and $(\nu^{C}_{R}(x), \nu^{C}_{L}(x))$, respectively, correspond to coordinates and momenta.
We also note that kinetic terms are transformed to kinetic terms, and mass terms are transformed to mass terms,
respectively, by the present linear transformation of fields. We understand this general unitary transformation $U$, which is readily defined for an arbitrary $n$ number of generations with $U(2n)$, as a generalization of the Pauli transformation~\cite{pauli} (originally defined for a single generation with $U(2)$) and we define C, P and CP at each frame of the Pauli transformation by looking at the kinetic part~\footnote{We define the term ``Pauli frame'' analogously to the term Lorentz frame by treating the transformation $U$ as an analogue of the Lorentz transformation.}. In contrast, the specific  transformation $U$ used in the diagonalization of mass terms \eqref{Autonne-Takagi} (Autonne-Takagi factorization)  is understood as a generalization of the Gursey transformation~\cite{gursey}. See \cite{schechter,balantekin} for the past discussions of related subjects.

A salient feature of our proposal is to define C and P at each Pauli frame specified by the generalized Pauli-Gursey transformation, namely, one Pauli frame with well-defined C and P is transformed to another Pauli frame with another set of well-defined C and P.  Note that the generalized Pauli-Gursey transformation mixes particles and antiparticles and thus  changes the definition of the vacuum. The CP symmetry is also generally modified by generalized Pauli-Gursey transformations; CP transformations in \eqref{C-P-CP1} and \eqref{C-P-CP2}, respectively,  imply 
\begin{eqnarray}\label{CP-relation}           
            \left(\begin{array}{c}
            \nu_{L}^{C}\\
            \nu_{L}
            \end{array}\right)
            = U \left(\begin{array}{c}
            \tilde{\nu}_{L}^{C}\\
            \tilde{\nu}_{L}
            \end{array}\right)
           \rightarrow
            i\gamma^{0}\left(\begin{array}{c}
            \nu_{R}\\
            \nu_{R}^{C}
            \end{array}\right)
            = i\gamma^{0} U 
            \left(\begin{array}{c}
            \tilde{\nu}_{R}\\
            \tilde{\nu}_{R}^{C}
            \end{array}\right)          
\end{eqnarray}
which does not agree with the second relation in \eqref{variable-change-0-prime} for $U \neq U^{\star}$ in general.

Since we know that a suitably chosen Pauli-Gursey transformation can diagonalize an arbitrary Lagrangian \eqref{Lagrangian} by a canonical transformation and lead to the  Lagrangian \eqref{exact-solution0}, one may attempt to define C, P and CP for the Lagrangian \eqref{exact-solution0} using the rules \eqref{C-P-CP2}. However, one can confirm that those C and P are not separately the symmetries of this total Lagrangian although CP is a good symmetry. See also \eqref{C-P-CP3} later.  To see the C-breaking explicitly,  one may define  Dirac-type variables 
\begin{eqnarray}\label{Dirac-type}
\tilde{\nu}(x)=\tilde{\nu}_{L}+\tilde{\nu}_{R}, \ \ \tilde{\nu}^{C}(x)=\tilde{\nu}^{C}_{L}+\tilde{\nu}^{C}_{R},
\end{eqnarray}
which are invertible by noting
$\tilde{\nu}_{L}=[(1-\gamma_{5})/2]\tilde{\nu}$ and  $\tilde{\nu}_{R}=[(1+\gamma_{5})/2]\tilde{\nu}$. Eq.\eqref{exact-solution0} or equivalently \eqref{exact-solution1} is then written as 
\begin{eqnarray}\label{exact-solution3}
{\cal L}
&=&\overline{\tilde{\nu}}(x)i\delslash \tilde{\nu}(x)
+(1/4)[\overline{\tilde{\nu}}(M_{1}+M_{2})\gamma_{5}\tilde{\nu}^{C}-
\overline{\tilde{\nu}^{C}}(M_{1}+M_{2})\gamma_{5}\tilde{\nu}]  \nonumber\\
&& -(1/4)[\overline{\tilde{\nu}}(M_{1}-M_{2})\tilde{\nu}^{C}+
\overline{\tilde{\nu}^{C}}(M_{1}-M_{2})\tilde{\nu}].
\end{eqnarray}
The terms with $(M_{1}-M_{2})$ are C-invariant under $\tilde{\nu}(x)\leftrightarrow
\tilde{\nu}^{C}(x)$, while the terms  with $(M_{1}+M_{2})$ are C-violating and correspond to a "C-condensation". The Lagrangian \eqref{exact-solution3} is an analogue of the BCS theory. 

Incidentally, mass terms are not generally invariant under generalized Pauli-Gursey transformations,  but it is natural to regard those mass terms, which are related by the action of $U(6)$, as belonging to an equivalence
class of mass terms.  

Our next strategy is to search for a suitable Pauli frame where C, P and CP defined by the kinetic part of the Lagrangian also become the good symmetries of the total Lagrangian.
We thus consider a further $6\times 6$ real generalized Pauli-Gursey transformation $O$, which is orthogonal and thus preserves CP according to \eqref{CP-relation}, by 
\begin{eqnarray} \label{orthogonal-variable-change}          
            &&\left(\begin{array}{c}
            \tilde{\nu}_{L}^{C}\\
            \tilde{\nu}_{L}           
            \end{array}\right)
            = O \left(\begin{array}{c}
            N_{L}^{C}\\
            N_{L}
            \end{array}\right)
           ,\ \ \ \ 
            \left(\begin{array}{c}
           \tilde{\nu}_{R}\\
            \tilde{\nu}_{R}^{C}
            \end{array}\right)
            = O
            \left(\begin{array}{c}
            N_{R}\\
            N_{R}^{C}
            \end{array}\right),          
\end{eqnarray}
we then obtain from \eqref{exact-solution0}
\begin{eqnarray}\label{BCS-like}
{\cal L}
&=&(1/2)\{\overline{N_{L}}(x)i\delslash N_{L}(x)+ \overline{N_{L}^{C}}(x)i\delslash N_{L}^{C}(x)\nonumber\\
&&+\overline{N_{R}}(x)i\delslash N_{R}(x) + \overline{N_{R}^{C}}(x)i\delslash N_{R}^{C}(x)\}\nonumber\\
&-&(1/2)\left(\begin{array}{cc}
            \overline{N_{R}},&\overline{N_{R}^{C}}
            \end{array}\right)
            O^{T}\left(\begin{array}{cc}
            M_{1}&0\\
            0&-M_{2} 
            \end{array}\right)O            
            \left(\begin{array}{c}
            N_{L}^{C}\\
            N_{L}
            \end{array}\right) +h.c..
\end{eqnarray}
We choose a specific $6\times6$ orthogonal transformation
\begin{eqnarray}\label{orthogonal1}
O=\frac{1}{\sqrt{2}}\left(\begin{array}{cc}
            1&1\\
            -1&1
            \end{array}\right)
\end{eqnarray}
where $1$ stands for a $3\times3$ unit matrix and then 
\begin{eqnarray}
O^{T}\left(\begin{array}{cc}
            M_{1}&0\\
            0&-M_{2} 
            \end{array}\right)O =\frac{1}{2}\left(\begin{array}{cc}
            M_{1}-M_{2}&M_{1}+M_{2}\\
            M_{1}+M_{2}&M_{1}-M_{2}
            \end{array}\right).
\end{eqnarray}
We thus have 
\begin{eqnarray}\label{BCS-like2}
{\cal L}
&=&(1/2)\{\overline{N}(x)i\delslash N(x)+ \overline{N^{C}}(x)i\delslash N^{C}(x)\}\nonumber\\
&-&(1/4)\{\overline{N}(M_{1}+M_{2})N+ \overline{N^{C}}(M_{1}+M_{2})N^{C}\}\nonumber\\
&-&(1/4)[\overline{N}(M_{1}-M_{2})N^{C}+\overline{N^{C}}(M_{1}-M_{2})N]
\end{eqnarray}
which is invariant under the C, P and CP defined by the kinetic part of the Lagrangian,
\begin{eqnarray}\label{C-P-CP-of-N}
C&:& N(x) \leftrightarrow N^{C}(x)=C\overline{N}^{T}(x),\nonumber\\
P&:& N(x) \rightarrow i\gamma^{0}N(t,-\vec{x}), \ \ N^{C}(x) \rightarrow i\gamma^{0}N^{C}(t,-\vec{x}),\nonumber\\
CP&:& N(x) \rightarrow i\gamma^{0}N^{C}(t,-\vec{x}), \ \ N^{C}(x) \rightarrow i\gamma^{0}N(t,-\vec{x}).
\end{eqnarray}
This is also the essence of a relativistic analogue of the Bogoliubov's canonical transformation, which transforms the C-violating terms with $(M_{1}+M_{2})$ in the notation of 
\eqref{exact-solution3} to the Dirac-type mass terms~\cite{FT,FT2}; in fact, the specific choice of the transformation  \eqref{orthogonal1} was made from this consideration of Bogoliubov transformation. Note that only the Dirac-type particles $N(x)$ and $N^{C}(x)$ with well-defined C, P and CP properties appear in this specific Pauli frame where the Lagrangian becomes C and thus P invariant and the chiral structure disappears.

We now make a renaming of variables 
\begin{eqnarray} \label{Majorana-variables}          
            \psi_{+}(x)=\frac{1}{\sqrt{2}}(N(x)+N^{C}(x)), \ \ 
            \psi_{-}(x)=\frac{1}{\sqrt{2}}(N(x)-N^{C}(x)),           
\end{eqnarray}
and we obtain
\begin{eqnarray}\label{Majorana1}
{\cal L}
&=&(1/2)\{\overline{\psi_{+}}(x)i\delslash \psi_{+}(x)+ \overline{\psi_{-}}(x)i\delslash \psi_{-}(x)\}\nonumber\\
&-&(1/2)\{\overline{\psi_{+}}M_{1}\psi_{+}+ \overline{\psi_{-}}M_{2}\psi_{-}\}.
\end{eqnarray}
After this renaming of variables, we find the transformation laws of $\psi_{\pm}(x)$ induced by those of $N$ and $N^{C}$ in \eqref{C-P-CP-of-N},
\begin{eqnarray}\label{final-C-P-CP}
C&:& \psi_{+}(x) \rightarrow \psi_{+}(x), \ \ \psi_{-}(x) \rightarrow - \psi_{-}(x),\nonumber\\
P&:& \psi_{+} \rightarrow i\gamma^{0}\psi_{+}(t,-\vec{x}), \ \ \psi_{-}(x) \rightarrow i\gamma^{0}\psi_{-}(t,-\vec{x}),\nonumber\\
CP&:& \psi_{+}(x) \rightarrow i\gamma^{0}\psi_{+}(t,-\vec{x}), \ \ \psi_{-}(x) \rightarrow -i\gamma^{0}\psi_{-}(t,-\vec{x})
\end{eqnarray}
which naturally keep the Lagrangian \eqref{Majorana1} invariant.
 When one defines a unitary charge conjugation operator ${\cal C}_{N}N(x){\cal C}^{\dagger}_{N}=N^{C}(x)=C\overline{N}^{T}(x)$, the operator ${\cal C}_{M}={\cal C}_{N}$ gives rise to ${\cal C}_{M}\psi_{\pm}(x){\cal C}^{\dagger}_{M}=\pm \psi_{\pm}(x)$. 
We thus determine {\em six} Majorana fermions $\psi_{\pm}(x)$ (each contains 3 flavor freedom) by the generalized Pauli-Gursey transformation~\footnote{The definition $\psi_{M}=\frac{1}{\sqrt{2}i}(N(x)-N^{C}(x))$ with an imaginary factor $i$ which satisfies classical relation $\psi_{M}=C\overline{\psi_{M}}^{T}$ is often used instead of our $\psi_{-}(x)=\frac{1}{\sqrt{2}}(N(x)-N^{C}(x))$, but this definition requires an anti-unitary ${\cal C}$ to maintain ${\cal C}\psi_{M}{\cal C}^{\dagger}=\psi_{M}$. }.

This analysis shows that one can define Majorana fermions in a natural manner by a suitable choice of Pauli frame, where C, P and CP defined by the kinetic part of the Lagrangian become the good symmetries of the total Lagrangian. The C symmetry to define Majorana neutrinos agrees with the C symmetry for $N(x)$ and $N^{C}(x)$. This definition of Majorana neutrinos agrees with the idea of Majorana neutrinos as Bogoliubov quasi-particles \cite{FT2}. 
We emphasize that the renaming \eqref{Majorana-variables} is consistent in the sense that the C transformation properties of $N(x)$ implied by the Lagrangian \eqref{BCS-like2} agree with the C transformation properties of $\psi_{\pm}(x)$ implied by the Lagrangian \eqref{Majorana1}.

In the definition of the above classical transformation rules of P and CP, we adopted  the  transformation rule of ``$i\gamma^{0}$ parity'' which is defined by
\begin{eqnarray}\label{i-gamma-P}
\psi_{L,R}^{P}(t,\vec{x})=i\gamma^{0}\psi_{R,L}(t,-\vec{x}).
\end{eqnarray}
The non-trivial phase freedom of the parity transformation in fermion number non-conserving theory has been analyzed by Weinberg \cite{weinberg1}. This definition of parity operation is the natural choice in a theory with Majorana fermions. The reason is that a  Majorana fermion $\psi_M(x)$, which satisfies $\psi_{M}(x)=C\overline{\psi_{M}}^{T}(x)$ classically,  stays Majorana after parity transformation,  i.e. the parity transformation preserves the Majorana condition: $C\overline{i\gamma^{0}\psi_M(t,-\vec{x})} = i\gamma^{0}C\overline{\psi_M(t,-\vec{x})}=i\gamma^{0}\psi_M(t,-\vec{x})$. This "$i\gamma^{0}$ parity" is crucial to assign a consistent parity to an isolated Majorana fermion~\cite{FT}~\footnote{
In the full theory with charged leptons included, we assign the $i\gamma^{0}$-parity to charged leptons, for example, $e(x)\rightarrow i\gamma^{0}e(t,-\vec{x})$ for the sake of consistency, although the extra phase is cancelled in the lepton-number conserving terms.}.

\section{Conventional seesaw formulation}
We comment on the conventional definition of Majorana neutrinos in the seesaw model from the point of view of the generalized Pauli-Gursey transformation.
One may attempt a direct renaming of variables defined by
\begin{eqnarray}  \label{Majorana-variables2}          
            &&\left(\begin{array}{c}
            \psi_{+L}(x)\\
            \psi_{-L}(x)
            \end{array}\right)=\left(\begin{array}{c}
            \tilde{\nu}_{L}^{C}\\
            \tilde{\nu}_{L}           
            \end{array}\right)
           ,\ \ \ \ 
            \left(\begin{array}{c}
            \psi_{+R}(x)\\
            \psi_{-R}(x)
            \end{array}\right)=\left(\begin{array}{c}
           \tilde{\nu}_{R}\\
            -\tilde{\nu}_{R}^{C}
            \end{array}\right),          
\end{eqnarray}
or equivalently
\begin{eqnarray}\label{Majorana-variables3}
\psi_{+}(x)=\tilde{\nu}_{R}+ \tilde{\nu}_{L}^{C}, \ \ 
\psi_{-}(x)=\tilde{\nu}_{L}- \tilde{\nu}_{R}^{C}
\end{eqnarray}
in the Lagrangian \eqref{exact-solution1}, and 
one obtains
\begin{eqnarray}\label{Majorana2}
{\cal L}
&=&(1/2)\{\overline{\psi_{+}}(x)i\delslash \psi_{+}(x)+ \overline{\psi_{-}}(x)i\delslash \psi_{-}(x)\}\nonumber\\
&-&(1/2)\{\overline{\psi_{+}}M_{1}\psi_{+}+ \overline{\psi_{-}}M_{2}\psi_{-}\}
\end{eqnarray}
which formally agrees with \eqref{Majorana1}. This is the common procedure in the seesaw mechanism~\cite{xing, mohapatra1, fukugita, giunti, bilenky,valle}.

The fields $\psi_{\pm}$ defined in \eqref{Majorana-variables3} are later shown to agree with the fields $\psi_{\pm}(x)=\frac{1}{\sqrt{2}}(N(x)\pm N^{C}(x))$ defined in \eqref{Majorana-variables} with certain qualifications, but let's pretend that they are independent for the moment.
The transformation rules of renamed variables $\psi_{\pm}$ induced by the original variables $\tilde{\nu}_{R}$ and $\tilde{\nu}_{L}$ in \eqref{C-P-CP2} are given by, in analogy with \eqref{final-C-P-CP},
\begin{eqnarray}\label{C-P-CP3}
C&:& \psi_{+R}(x) \leftrightarrow - \psi_{-R}(x),\ \ \psi_{+L}(x) \leftrightarrow  \psi_{-L}(x),\nonumber\\
P&:& \psi_{+R}(x) \rightarrow i\gamma^{0}\psi_{-L}(t,-\vec{x}), \ \ \psi_{+L}(x) \rightarrow -i\gamma^{0}\psi_{-R}(t,-\vec{x}),\nonumber\\
&& \psi_{-L}(x) \rightarrow i\gamma^{0}\psi_{+R}(t,-\vec{x}), \ \ \psi_{-R}(x) \rightarrow -i\gamma^{0}\psi_{+L}(t,-\vec{x}),
\nonumber\\
CP&:& \psi_{+}(x) \rightarrow i\gamma^{0}\psi_{+}(t,-\vec{x}), \ \ \psi_{-}(x) \rightarrow -i\gamma^{0}\psi_{-}(t,-\vec{x}).
\end{eqnarray}
One can confirm that these symmetry transformation rules keep the kinetic part of the Lagrangian \eqref{Majorana2}
invariant, but the total Lagrangian is not invariant. 

One may however recognize that the variables in \eqref{Majorana-variables3} satisfy the classical Majorana conditions
\begin{eqnarray} \label{identities}          
&&C\overline{\psi_{+}(x)}^{T}=C\overline{\tilde{\nu}_{R}}^{T}(x)+C\overline{C\overline{\tilde{\nu}_{R}}^{T}}^{T}(x)=C\overline{\tilde{\nu}_{R}}^{T}(x)+\tilde{\nu}_{R}(x)=\psi_{+}(x),\nonumber\\
&&C\overline{\psi_{-}(x)}^{T}=C\overline{\tilde{\nu}_{L}}^{T}(x)-C\overline{C\overline{\tilde{\nu}_{L}}^{T}}^{T}(x)=C\overline{\tilde{\nu}_{L}}^{T}(x)-\tilde{\nu}_{L}(x)=-\psi_{-}(x)
\end{eqnarray}
if one recalls  $\tilde{\nu}_{L}^{C}(x)=C\overline{\tilde{\nu}_{R}}^{T}(x)$ and $\tilde{\nu}^{C}_{R}(x)=C\overline{\tilde{\nu}_{L}}^{T}(x)$ in \eqref{C-P-CP2}.
One may thus assume the existence of some generic operator ${\cal C}$ which satisfies 
\begin{eqnarray} \label{pseudo C}          
&&{\cal C}\psi_{+}(x){\cal C}^{\dagger}={\cal C}\tilde{\nu}_{R}(x){\cal C}^{\dagger}+{\cal C}C\overline{\tilde{\nu}_{R}}^{T}(x){\cal C}^{\dagger}=C\overline{\tilde{\nu}_{R}}^{T}(x)+\tilde{\nu}_{R}(x)=\psi_{+}(x),\nonumber\\
&&{\cal C}\psi_{-}(x){\cal C}^{\dagger}={\cal C}\tilde{\nu}_{L}(x){\cal C}^{\dagger}-{\cal C}C\overline{\tilde{\nu}_{L}}^{T}(x){\cal C}^{\dagger}=C\overline{\tilde{\nu}_{L}}^{T}(x)-\tilde{\nu}_{L}(x)=-\psi_{-}(x).
\end{eqnarray}
From the comparison of \eqref{identities} and \eqref{pseudo C}, it may appear natural to guess that the operator ${\cal C}$ acts as follows \cite{xing, fukugita, giunti, bilenky}:
\begin{eqnarray} \label{pseudo C operator}          
&&{\cal C}\tilde{\nu}_{R}(x){\cal C}^{\dagger}=C\overline{\tilde{\nu}_{R}}^{T}(x),
\ \ \ {\cal C}C\overline{\tilde{\nu}_{R}}^{T}(x){\cal C}^{\dagger}=\tilde{\nu}_{R}(x),\nonumber\\
&&{\cal C}\tilde{\nu}_{L}(x){\cal C}^{\dagger}=C\overline{\tilde{\nu}_{L}}^{T}(x), \ \ \ {\cal C}C\overline{\tilde{\nu}_{L}}^{T}(x){\cal C}^{\dagger}=\tilde{\nu}_{L}(x).
\end{eqnarray}
However, if one should assume the existence of a well-defined unitary operator ${\cal C}\tilde{\nu}_{R}(x){\cal C}^{\dagger}=C\overline{\tilde{\nu}_{R}}^{T}(x)$ as in \eqref{pseudo C operator}, for example, one would encounter an inconsistency 
\begin{eqnarray}\label{operatorial inconsistency}
{\cal C}\tilde{\nu}_{R}(x) {\cal C}^{\dagger}=\frac{1}{2}(1+\gamma_{5})
{\cal C}\tilde{\nu}_{R}(x){\cal C}^{\dagger}
=\frac{1}{2}(1+\gamma_{5})C\overline{\tilde{\nu}_{R}}^{T}(x)=0
\end{eqnarray}
 by noting that $\tilde{\nu}_{R}(x)
=\frac{1}{2}(1+\gamma_{5})\tilde{\nu}_{R}(x)
$ and $C\overline{\tilde{\nu}_{R}}^{T}(x)$ is left-handed~\cite{FT,FT2}.        
It has been recently shown that the symmetry \eqref{pseudo C operator} is in fact a hidden symmetry associated with CP symmetry but operatorially undefined \cite{FT3}; the inconsistency \eqref{operatorial inconsistency} is an example of such operatorial indefiniteness.  For this reason, the naming {\em pseudo C-symmetry} was suggested for \eqref{pseudo C operator} in \cite{FT3}.

To understand the origin of the operatorial difficulty of the pseudo C-symmetry \eqref{operatorial inconsistency} from the present point of view, it is important to recall that one cannot simultaneously accommodate the standard C transformation laws for chiral components \eqref{C-P-CP2} as indicated by \eqref{C-P-CP3} and the C transformation laws expected for Majorana fermions, $\psi_{\pm}(x) \rightarrow \pm \psi_{\pm}(x)$, indicated by \eqref{identities} at an {\em identical} Pauli frame (except for the very special case \eqref{Majorana-variables}). One may recall that, in our proposed Pauli-Gursey transformation, the C transformation rules are all standard ones at any frame, in a generic notation, $\nu_{L}(x)\rightarrow C\overline{\nu_{R}}^{T}(x)$ and $\nu_{R}(x)\rightarrow C\overline{\nu_{L}}^{T}(x)$.\footnote{The standard C transformation laws of $N_{R}(x)$ and $N_{L}(x)$ give rise to the Majorana transformation laws of $\psi_{\pm}(x)$ in \eqref{final-C-P-CP}.} The transformation rule 
\eqref{pseudo C}  is regarded as assuming a well defined C transformation law 
to both $\nu_{L,R}(x)$ and $\psi_{\pm}(x)$ at the same Pauli frame, which may give an intuitive understanding of the difficulty \eqref{operatorial inconsistency}. 

We would next like to discuss the more general issue, namely, how the C and P violating Lagrangian 
\eqref{exact-solution1} or equivalently \eqref{exact-solution3} can be consistent with the Majorana solutions with good C, P and CP symmetries. This analysis also helps understand the above difficulty \eqref{operatorial inconsistency} better. For a clear analysis of this issue, we use the notation of quantized theory. 
One can confirm that the relations \eqref{Majorana-variables2} are also valid for the variables in  \eqref{orthogonal-variable-change} and \eqref{Majorana-variables}, but \eqref{Majorana-variables2} are now re-written in the form 
\begin{eqnarray}\label{Weyl-Majorana-conversion}
             \left(\begin{array}{c}
            \tilde{\nu}_{L}(x)\\
            C\overline{\tilde{\nu}_{R}}^{T}(x)
            \end{array}\right)
            &=&\left(\begin{array}{c}
            \frac{1-\gamma_{5}}{2}      
            \psi_{-}(x)\\
            \frac{1-\gamma_{5}}{2}\psi_{+}(x)            
            \end{array}\right),\nonumber\\
 \left(\begin{array}{c}
            \tilde{\nu}_{R}(x)\\
            C\overline{\tilde{\nu}_{L}}^{T}(x)
            \end{array}\right)
            &=&\left(\begin{array}{c}
            \frac{1+\gamma_{5}}{2}\psi_{+}(x)\\
            -\frac{1+\gamma_{5}}{2}\psi_{-}(x)            
            \end{array}\right).           
\end{eqnarray}
In these two relations \eqref{Majorana-variables2} and \eqref{Weyl-Majorana-conversion}, we regard the variables appearing on the right-hand sides are the variables defined in their own Pauli frames. We now restate the defining transformation rules of $\tilde{\nu}_{L,R}(x)$ in \eqref{C-P-CP2}
\begin{eqnarray}\label{C-P-CP2-prime}
C&:& \tilde{\nu}_{L,R}(x) \rightarrow \tilde{\nu}^{C}_{L,R}(x) = C\overline{\tilde{\nu}_{R,L}}^{T}(x),\nonumber\\
P&:& \tilde{\nu}_{L,R}(x) \rightarrow i\gamma^{0}\tilde{\nu}_{R,L}(t,-\vec{x}),\ \ \tilde{\nu}^{C}_{L,R}(x) \rightarrow i\gamma^{0}\tilde{\nu}^{C}_{R,L}(t,-\vec{x}),\nonumber\\
CP&:& \tilde{\nu}_{L,R}(x)\rightarrow i\gamma^{0}C\overline{\tilde{\nu}_{L,R}(t,-\vec{x})}^{T}
\end{eqnarray}
which give rise to \eqref{C-P-CP3}.
On the other hand, the variables $\psi_\pm(x)$ in \eqref{Weyl-Majorana-conversion} satisfy the Majorana relations
\begin{eqnarray}\label{quantum C}
&&\psi^C_\pm(x)\equiv {\cal C}_M\psi_\pm(x){\cal C}_M^{\dagger}=C\overline\psi_\pm^T(x)=\pm \psi_\pm(x),\nonumber\\
&&{\cal P}_{M}\psi_{\pm}(x) {\cal P}^{\dagger}_{M}=i\gamma^{0}\psi_{\pm}(t,{-\vec{x}}), \nonumber\\
&& ({\cal C}_M{\cal P}_{M})\psi_{\pm}(x)({\cal C}_M{\cal P}_{M})^{\dagger}= \pm i\gamma^{0}\psi_{\pm}(t,{-\vec{x}})
\end{eqnarray}
defined by \eqref{final-C-P-CP} at the natural Pauli frame. 
Using the renaming of variables in \eqref{Weyl-Majorana-conversion}, one obtains the transformation laws
\begin{eqnarray}\label{induced-transformations}
&&{\cal C}_{M}\tilde{\nu}_{L}(x){\cal C}^{\dagger}_{M}= -\tilde{\nu}_{L}(x),  \ \
{\cal C}_{M}\tilde{\nu}_{R}(x){\cal C}^{\dagger}_{M}= \tilde{\nu}_{R}(x),
\nonumber\\
&&{\cal P}_{M}\tilde{\nu}_{L}(x){\cal P}^{\dagger}_{M}= -i\gamma^{0}C\overline{\tilde{\nu}_{L}}^{T}(t,-\vec{x}),\ \ {\cal P}_{M}\tilde{\nu}_{R}(x){\cal P}^{\dagger}_{M}= i\gamma^{0}C\overline{\tilde{\nu}_{R}}^{T}(t,-\vec{x}),\nonumber\\
&&({\cal CP})_{M} \tilde{\nu}_{L,R}(x)({\cal CP})^{\dagger}_{M}= i\gamma^{0}C\overline{\tilde{\nu}_{L,R}(t,-\vec{x})}^{T}.
\end{eqnarray}
In the generalized Pauli-Gursey transformation, the C transformation rules are all standard ones at any frame, in a generic notation,  $\nu_{L}(x)\rightarrow C\overline{\nu_{R}}^{T}(x)$ and $\nu_{R}(x)\rightarrow C\overline{\nu_{L}}^{T}(x)$ (doublet representations) as in \eqref{C-P-CP2-prime}. But when one looks back at how one arrived at a specific frame which defines the Majorana fermions naturally, as in \eqref{C-P-CP-of-N}, the natural C transformation laws look very different in terms of the  variables in the old frame.  In \eqref{induced-transformations}, we thus specified the Pauli frame explicitly where the operators C, P and CP are defined, and thus our analysis relies on the notation of quantized theory. This is how unorthodox C transformation (singlet representations) and parity transformation laws in \eqref{induced-transformations}, which are very different from those in \eqref{C-P-CP2-prime},  are recognized when looked back from the natural Pauli frame of Majorana fermions. 

From the above analysis, we see that the specific 
(real orthogonal) Pauli-Gursey transformation \eqref{orthogonal-variable-change} illustrates a mechanism how the Majorana neutrinos in the conventional formulation are defined  in the seesaw Lagrangian \eqref{exact-solution1}, which does not appear to be C symmetric, if one adopts effectively the new (induced) C symmetry \eqref{induced-transformations} for chiral fermions that renders the Lagrangian \eqref{exact-solution1} C-invariant.
A salient feature of this mechanism of the induced new C symmetry is the inevitable appearance of a singlet (trivial) representation of C symmetry, such as $\nu_{R}(x)\rightarrow \nu_{R}(x)$ and $\nu_{L}(x)\rightarrow -\nu_{L}(x)$,  for chiral fermions; the trivial representation renders any Lagrangian invariant.
 
Quite generally, if one has the identical good CP symmetry as in \eqref{C-P-CP2-prime} and 
\eqref{induced-transformations}, one may re-define C to be trivial in the C-violating theory \eqref{C-P-CP2-prime} and P to be identical to CP, 
one would then formally arrive at the representation with good C, P and CP symmetries \eqref{induced-transformations}~\footnote{We usually assign only the well-defined CP, $\nu_{L}(x)\rightarrow i\gamma^{0}C\overline{\nu_{L}(t,-\vec{x})}^{T}$, to the Weyl fermion ${\cal L}=\overline{\nu_{L}}(x)i\delslash \nu_{L}(x)$. But one may formally assign C and P also if one assumes a trivial representation of C and thus P which is identical to CP.  Weinberg's model of Majorana neutrinos~\cite{weinberg2} 
\begin{eqnarray}\label{weinberg-type}
{\cal L}&=&\overline{\nu_{L}}(x)i\delslash \nu_{L}(x)
-(1/2)\{\nu_{L}^{T}(x)Cm_{L}\nu_{L}(x) + h.c.\}
=(1/2)\{\overline{\psi}(x)i\delslash \psi(x)- \overline{\psi}(x)m_{L}\psi(x)\}\nonumber           
\end{eqnarray}
with $\psi(x)=\nu_{L}(x)+ C\overline{\nu_{L}}^{T}(x)$ could formally be understood in this spirit also, after diagonalizing the symmetric complex mass matrix (and thus removing possible CP breaking) by the $3 \times 3$ Autonne-Takagi transformation $\nu_{L}\rightarrow U\nu_{L}$. The field $\psi(x)$ satisfies ${\cal C}\psi(x){\cal C}^{\dagger}= \psi(x)$ and $({\cal CP})\psi(x)({\cal CP})^{\dagger}={\cal P}\psi(x){\cal P}^{\dagger}= i\gamma^{0}\psi(t,-\vec{x})$ in this interpretation. }.

Finally, the C transformation laws in \eqref{C-P-CP2-prime} and 
\eqref{induced-transformations} are very different but CP transformation laws are identical, as we already noted,  since the orthogonal transformation \eqref{orthogonal-variable-change} does not modify CP.
The pseudo C-symmetry \eqref{pseudo C operator} often used in the seesaw mechanism~\cite{xing, fukugita, giunti, bilenky}  may be regarded as being extracted from these common CP transformation laws. In fact it has been recently shown  that the pseudo C-symmetry is a hidden symmetry associated with CP but operatorially undefined \cite{FT3}.

\section{Conclusion}
The generalized Pauli-Gursey transformation provides a powerful machinery to analyze the seesaw mechanism in a systematic way, and the present analysis may support a view~\cite{FT2} that the Majorana neutrinos as Bogoliubov quasi-particles defined by C, as in \eqref{C-P-CP-of-N}, \eqref{Majorana-variables} and \eqref{final-C-P-CP}, is the most natural picture of neutrinos in the seesaw model. The generalized Pauli-Gursey transformation which changes the definition of the vacuum  shows that the definitions of C and P strongly depend on the Pauli frame we choose and a proper use of  CP transformation $\nu_{L,R}(x)\rightarrow i\gamma^{0}C\overline{\nu_{L,R}(t,-\vec{x})}^{T}$,  which is better specified than C, may open a path to an alternative treatment of Majorana neutrinos in the conventional formulation of seesaw mechanism.
\\

I thank Anca Tureanu for very helpful discussions. I also thank the hospitality of Helsinki Institute of Physics, University of Helsinki.  
This work is supported in part by JSPS KAKENHI (Grant No.18K03633).

\end{document}